\documentclass[12pt,a4paper]{article}
\usepackage{axodraw}
\usepackage{graphicx}
\newcommand{\vs}{\vspace{-0.25cm}}

\begin{document}

\begin{center}

{\Large
\textbf{Nuclear energy density functional from\\ chiral pion-nucleon 
dynamics revisited}\footnote{Work supported in part by BMBF, GSI and the DFG
cluster of excellence: Origin and Structure of the Universe.}} 

\bigskip
N. Kaiser and W. Weise\\

\bigskip

{\small Physik Department, Technische Universit\"{a}t M\"{u}nchen, 
D-85747 Garching, Germany\\

\smallskip

{\it email: nkaiser@ph.tum.de}}

\end{center}

\bigskip

\begin{abstract}
We use a recently improved density-matrix expansion to calculate the nuclear 
energy density functional in the framework of in-medium chiral perturbation 
theory. Our calculation treats systematically the effects from $1\pi$-exchange,
iterated $1\pi$-exchange, and irreducible $2\pi$-exchange with intermediate 
$\Delta$-isobar excitations, including Pauli-blocking corrections up to 
three-loop order. We find that the effective nucleon mass $M^*(\rho)$ 
entering the energy density functional is identical to the one of Fermi-liquid 
theory when employing the improved density-matrix expansion. The strength 
$F_\nabla(\rho)$ of the $(\vec\nabla \rho)^2$ surface-term as provided by 
the pion-exchange dynamics is in good agreement with that of phenomenological
Skyrme forces in the density region $\rho_0/2 <\rho < \rho_0$. The spin-orbit
coupling strength $F_{so}(\rho)$ receives contributions from iterated
$1\pi$-exchange (of the ``wrong sign'') and from three-nucleon interactions
mediated by $2\pi$-exchange with virtual $\Delta$-excitation (of the ``correct
sign''). In the region around $\rho_0/2 \simeq 0.08\,$fm$^{-3}$ where the
spin-orbit interaction in nuclei gains most of its weight these two components 
tend to cancel, thus leaving all room for the short-range spin-orbit 
interaction.  The strength function $F_J(\rho)$ multiplying the square of the 
spin-orbit density comes out much larger than in phenomenological Skyrme forces 
and it has a pronounced density dependence.

\end{abstract}

\bigskip


\bigskip 

\section{Introduction}
The nuclear energy density functional approach is the many-body method of 
choice in order to calculate the properties of medium-mass and heavy nuclei in 
a systematic manner \cite{reinhard}. In this context non-relativistic Skyrme 
forces \cite{skyrme,sk3,skmstar,sly,pearson} have gained much popularity 
because of their analytical simplicity and their ability to reproduce nuclear 
properties over the whole periodic table within the self-consistent 
Hartree-Fock approximation. Another widely and successfully used approach to 
nuclear  structure calculations are relativistic mean-field models 
\cite{walecka,ringreview}. In these models the nucleus is described as a 
collection of independent Dirac quasi-particles moving in self-consistently 
generated scalar and vector mean-fields. The footprints of relativity become 
visible  through the large nuclear spin-orbit interaction which emerges in 
that framework from the interplay of the strong scalar and vector mean-fields.
These counteract in producing the (attractive) central potential but act 
coherently to generate the strong spin-orbit potential. Constraints from chiral 
(pion-nucleon) dynamics and the symmetry breaking pattern of QCD at low 
energies have been implemented into a pertinent relativistic point-coupling 
Lagrangian in ref.\cite{finelli}. In such a more constrained scheme the 
results for nuclear observables come out comparable to those of the best 
phenomenological parameterizations of relativistic mean-field models.

A complementary approach in the quest for predictive nuclear energy density
functionals \cite{lesinski,drut,platter} focusses less on the fitting of 
experimental data, but attempts to constrain the analytical form of the 
functional and the values of its couplings from many-body perturbation theory
and the underlying two- and three-nucleon interactions. Switching from the
conventional hard-core NN-potentials to low-momentum interactions is essential
in this respect, because the nuclear many-body problem formulated in terms of 
low-momentum interactions becomes significantly more perturbative. Indeed, 
second-order perturbative calculations provide already a good account of the  
bulk correlations in infinite nuclear matter \cite{achim} and in doubly-magic 
nuclei \cite{roth}. 

In many-body perturbation theory the contributions to the energy are written 
in terms of density-matrices and propagators convoluted with the finite-range 
interaction vertices, and are therefore highly non-local in both space and 
time. In order to make such functionals numerically tractable in heavy 
open-shell nuclei it is desirable to develop simplified approximations to these 
functionals expressed in terms of local densities and currents only. At this 
stage of the procedure the density-matrix expansion comes prominently into 
play as it removes the non-local character of the exchange (Fock) contribution 
to the energy by mapping it onto a generalized Skyrme functional with density 
dependent couplings. Until recently, the prototype for that has been the 
density-matrix expansion of Negele and Vautherin \cite{negele}. This version 
of the density-matrix expansion (in particular its Fourier-transform to momentum
space) has also been used in refs.\cite{efun,deltamat} to calculate the
nuclear energy density functional in the framework of in-medium chiral 
perturbation theory. The density-dependent coupling strengths of the surface 
term, $(\vec \nabla \rho)^2$, or the spin-orbit term, $\vec \nabla \rho \cdot 
\vec J$, arise in these calculations exclusively from the long-range 
$1\pi$- and  $2\pi$-exchange dynamics in an inhomogeneous many-nucleon system 
characterized by a local density $\rho(\vec r\,)$ and a local spin-orbit 
density  $\vec J(\vec r\,)$.

In a recent paper by Gebremariam, Duguet and Bogner \cite{dmeimprov} an
improved density-matrix expansion has been developed for spin-unsaturated
nuclei. It has been demonstrated that phase-space averaging techniques allow
for a consistent expansion of both the spin-independent (scalar) part as well
as the spin-dependent (vector) part of the density-matrix. A further key 
feature of the new method has been to take into account the deformation 
displayed by the local density distribution at the surface of most nuclei. The 
accuracy of the new phase-space averaged density-matrix expansion and the 
original one of Negele and Vautherin has been gauged via the Fock energy 
(densities) arising from (schematic finite-range) central, tensor and 
spin-orbit interactions for a large set of semi-magic nuclei. For a central 
force the Fock energy depends primarily on the spin-independent (scalar) 
part of the  density-matrix and a few percent  accuracy is reached for both 
variants of the density-matrix expansion. On the other hand the Fock energy 
due to a tensor force is determined by the spin-dependent (vector) part of 
the density-matrix. In that case the original density-matrix expansion of 
Negele and Vautherin leads to an error of about $50\%$, whereas the new one 
based on phase-space averaging techniques reduces the error drastically to only 
a few percent. This is the same level of accuracy as obtained for interaction 
terms involving the spin-independent  (scalar) part of the density-matrix. For 
further details on these extensive and instructive test studies we refer to 
ref.\cite{dmeimprov}.

The purpose of the present work is to match with these new developments and to
reconsider the nuclear energy density functional as it emerges from chiral 
pion-nucleon dynamics on the basis of the improved density-matrix expansion of 
ref.\cite{dmeimprov}. Our paper is organized as follows. In section 2 we 
recall the explicit form of the improved density-matrix expansion of 
Gebremariam, Duguet and Bogner  \cite{dmeimprov}. Its Fourier-transform to 
momentum space provides the adequate technical tool to calculate the nuclear 
energy density functional in a diagrammatic framework. As a first 
interesting result we find that for the zero-range Skyrme force the new and the
old density-matrix expansion lead to identical results. Differences between
the two versions are therefore to be expected for the interaction contributions
arising from the long-range $1\pi$- and $2\pi$-exchange between nucleons. In
section 3, we present the analytical results for the density-dependent 
strength functions $F_\tau(\rho)$, $F_{so}(\rho)$ and $F_J(\rho)$ from which the 
nuclear energy density functional is composed. We restrict ourselves here to 
the isospin-symmetric case of equal proton and neutron number. These 
analytical expressions give individually the effects due to  $1\pi$-exchange, 
iterated $1\pi$-exchange, and irreducible $2\pi$-exchange with intermediate  
$\Delta$-isobar excitations, including Pauli-blocking corrections up to 
three-loop order. Section 4 is devoted to a discussion of our numerical 
results and finally section 5 ends with a summary and concluding remarks. 
In the appendix the three-body spin-orbit coupling strength  $F_{so}(\rho)$ is 
presented for an alternative description of the $2\pi$-exchange three-nucleon 
interaction.

\section{Improved density-matrix expansion and energy density 
functional}
The starting point for the construction of an explicit nuclear energy density
functional is the density-matrix as given by a sum over the occupied energy
eigenfunctions $\Psi_\alpha$ of the (non-relativistic) many-fermion system. 
According to Gebremariam, Duguet and Bogner \cite{dmeimprov} the bilocal 
density-matrix can be expanded in relative and center-of-mass coordinates, 
$\vec a$  and $\vec r$, as follows:\footnote{We are considering for equal 
proton and neutron number the spherical phase-space averaged version without 
quadrupolar deformation of the local Fermi momentum distribution. It brings 
about already most of the improvements \cite{dmeimprov}.}   
\begin{eqnarray} \sum_{\alpha}\Psi_\alpha( \vec r -\vec a/2)\Psi_\alpha^
\dagger(\vec r +\vec a/2) &=& {3 \rho\over a k_f}\, j_1(a k_f)-{a \over 2k_f} 
\,j_1(a k_f) \bigg[ \tau - {3\over 5} \rho k_f^2 - {1\over 4} \vec \nabla^2 
\rho \bigg] \nonumber \\ && + {3i \over 2a k_f} \,j_1(a k_f)\, \vec \sigma
\cdot (\vec a \times \vec J\,) + \dots\,,  \end{eqnarray}
where $j_1(x) = (\sin x - x \cos x)/x^2$ is the spherical Bessel function of
index 1. The other quantities appearing on the right hand side of eq.(1) are
the local nucleon density: 
\begin{equation} \rho(\vec r\,) =  {2k_f^3(\vec r\,)\over 3\pi^2} =  \sum_\alpha
\Psi^\dagger_\alpha( \vec r\,)\Psi_\alpha( \vec r\,)
\,,\end{equation} 
written here in terms of the local Fermi-momentum $k_f(\vec r\,)$, the local 
kinetic energy density: 
\begin{equation} \tau(\vec r\,) = \sum_{\alpha}\vec \nabla\Psi^\dagger
_\alpha( \vec r\,)\cdot \vec \nabla\Psi_\alpha( \vec r\,) \,,\end{equation}
and the local spin-orbit density:
\begin{equation} \vec J(\vec r\,) = \sum_{\alpha} \Psi^\dagger_\alpha( 
\vec r\,)i\,\vec \sigma \times \vec \nabla\Psi_\alpha( \vec r\,) \,.
\end{equation}
For notational simplicity we have dropped their argument $\vec r$ in eq.(1)
and will do so in the following. It is important to note that a pairwise
filling of time-reversed orbitals $\alpha$ has been assumed in eq.(1), so that 
(various possible) time-reversal-odd fields do not come into play 
\cite{reinhard}. The main difference of this improved density-matrix expansion 
to the original one of Negele and Vautherin \cite{negele} lies in the index of 
the Bessel function multiplying the kinetic energy and spin-orbit densities in
eq.(1). The Fourier-transform of the (expanded) density-matrix with respect to
both coordinates $\vec a$ and $\vec r$ defines a ''medium insertion'' for the
inhomogeneous many-nucleon system characterized by the time-reversal-even
fields $\rho(\vec r\,)$, $\tau(\vec r\,)$ and $\vec J(\vec r\,)$:        
\begin{eqnarray} \Gamma(\vec p,\vec q\,)& =& \int d^3 r \, e^{-i \vec q \cdot
\vec r}\,\bigg\{ \theta(k_f-|\vec p\,|) +{\pi^2 \over 4k_f^4}\Big[k_f\,\delta'
(k_f-|\vec p\,|)-2 \delta(k_f-|\vec p\,|)\Big] \nonumber \\ && \times \bigg( 
\tau - {3\over 5} \rho k_f^2 - {1\over 4} \vec \nabla^2 \rho \bigg) -{3\pi^2 
\over 4k_f^4}\,\delta(k_f-|\vec p\,|) \, \vec \sigma \cdot (\vec p \times 
\vec J\,)  \bigg\}\,.  \end{eqnarray}
The double line in the left picture of Fig.\,1 symbolizes this medium
insertion together with the assignment of the out- and in-going nucleon momenta
$\vec p \pm \vec q/2$. The momentum transfer $\vec q $ is provided by the
Fourier components of the inhomogeneous (matter) distributions 
$\rho(\vec r\,)$, $\tau(\vec r\,)$ and $\vec J(\vec r\,)$. As a check one
verifies that the Fourier transform $(1/2\pi^3)\int d^3 p\, e^{-i \vec p \cdot
\vec a}$ of the expression in the curly brackets in eq.(5) reproduces exactly 
the right hand side of the (improved) density-matrix expansion written in 
eq.(1). In comparison to the version of $\Gamma(\vec p,\vec q\,)$ which
followed from  Negele and Vautherin's density-matrix expansion \cite{efun} the 
weight function of the kinetic energy density $\tau(\vec r\,)$ has changed from 
$35(5\vec p^{\,2} -3k_f^2) \theta(k_f-|\vec p\,|) $ to $2k_f^3[k_f\,\delta'(k_f
-|\vec p\,|)-2 \delta(k_f-|\vec p\,|)]$ and that of the spin-orbit density
$\vec J(\vec r\,)$ has changed from  $\delta(k_f-|\vec p\,|) -k_f\,\delta'(k_f-
|\vec p\,|)$ to $-3 \delta(k_f-|\vec p\,|)$. For an inhomogeneous
many-nucleon system this leads to a different weighting of the momentum
dependent nucleon-nucleon interactions in the vicinity of the local 
Fermi-surface $|\vec p\,|=k_f(\vec r\,)$, with appropriate consequences for 
the energy density functional.

Going up to second order in spatial gradients (i.e. deviations from 
homogeneity) the energy density functional relevant for $N=Z$ even-even nuclei 
reads:  
\begin{eqnarray} {\cal E}[\rho,\tau,\vec J\,] &=& \rho\,\bar E(\rho)+\bigg[\tau-
{3\over 5} \rho k_f^2\bigg] \bigg[{1\over 2M}-{k_f^2 \over 4M^3}+F_\tau(\rho)
\bigg] \nonumber \\ && + (\vec \nabla \rho)^2\, F_\nabla(\rho)+  \vec \nabla 
\rho \cdot\vec J\, F_{so}(\rho)+ \vec J\,^2 \, F_J(\rho)\,.\end{eqnarray} 
Here, $\bar E(\rho)$ is the energy per particle of isospin-symmetric nuclear
matter evaluated at the local nucleon density $\rho(\vec r\,)$. The (small)
correction term $-k_f^2/4M^3$ in eq.(6) stems from the relativistically improved
kinetic energy and reflects in this way the relativistic increase of mass. The
density-dependent functions $F_\tau(\rho)$, $F_\nabla(\rho)$,  $F_{so}(\rho)$
and $F_J(\rho)$ arising from two- and three-nucleon interactions encode new
dynamical information specific for the inhomogeneous many-nucleon system. In
particular, $F_\nabla(\rho)$ measures the energy associated with density 
gradients at the nuclear surface and $F_\nabla(\rho)$ gives the strength of
the spin-orbit coupling. 

Returning to eq.(5) one sees that $F_\tau(\rho)$ emerges via a perturbation on
top of the density of states $\theta(k_f-|\vec p\,|)$. The single-particle
potential in nuclear matter can be obtained in the same way by introducing a
(three-dimensional) delta-function as the perturbation. Consequently, the 
strength function $F_\tau(\rho)$ can be expressed in terms of the momentum and 
density-dependent single-particle potential $U(p,k_f)$ as follows:    
\begin{equation} F_\tau(\rho) = {1 \over 2k_f} {\partial U(p,k_f)
    \over \partial p}\Big|_{p=k_f}\,.\end{equation}
In eq.(5) the term $\tau -3\rho k_f^2/5$ is accompanied by $-\vec\nabla^2\rho/4$.
Performing a partial integration of the energy $\int \!d^3 r\,{\cal E}[\rho,
\tau,\vec J\,]$ one is lead to the  decomposition:     
\begin{equation} F_\nabla(\rho) = {1\over 4}\, {\partial F_\tau(\rho)
\over  \partial \rho} +F_d(\rho) \,,\end{equation}
where $F_d(\rho)$ comprises all those contributions for which the $(\vec \nabla
\rho)^2$-factor originates directly from the momentum dependence of the 
interactions in an expansion up to order $\vec q^{\,2}$. Since no information
about the density-matrix expansion beyond its (fixed) nucleon matter piece 
$\theta(k_f-|\vec p\,|)$ goes into the derivation of the strength function 
$F_d(\rho)$ the pertinent contributions from the $2\pi$-exchange dynamics are 
still given in unchanged form by eqs.(12,15,19,24,28) in ref.\cite{efun} and 
eqs.(26,30) in 
ref.\cite{deltamat}. 
     
As a first test case for the improved density-matrix expansion (summarized in
eq.(5)) we have applied it to the (zero-range) Skyrme force \cite{skyrme,ring} 
and found that it gives identical results:
\begin{eqnarray} && F_\tau(\rho)^{(\rm Sk)} = {\rho \over 16}(3t_1+5t_2)\,, \qquad  
F_d(\rho)^{(\rm Sk)}  = {1\over 32}(3t_1-5t_2)\,,\nonumber \\ && F_{so}(\rho)^{(\rm Sk)} 
={3\over 4}W_0\,, \qquad  F_J(\rho)^{(\rm Sk)}={1\over 32}(t_1-t_2)\,,\end{eqnarray}
for the energy density functional as the original density-matrix expansion of 
Negele and Vautherin \cite{negele}. Obviously, for contact-interactions with 
their simple quadratic momentum dependence the different weighting of 
interaction strength in the vicinity of the Fermi-surface has no visible 
effect. A stronger influence of the actual form of the density-matrix 
expansion is therefore expected for the contributions arising from the 
long-range $1\pi$- and $2\pi$-exchange.  The pertinent analytical expressions 
are collected in the next section.      

\begin{figure}
\begin{center}
\includegraphics[scale=1.0,clip]{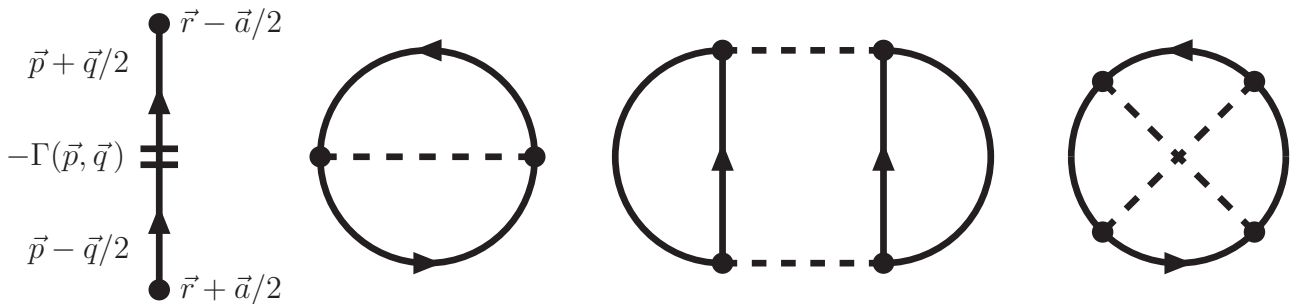}
\end{center}
\vspace{-0.2cm}
\caption{Left: The double line symbolizes the medium insertion defined by
eq.(5). Next are shown: The one-pion exchange Fock diagram and the iterated 
one-pion exchange Hartree and Fock diagrams. Their isospin factors for 
isospin-symmetric nuclear systems are 6, 12 and $-$6, respectively.}
\end{figure}

\section{Diagrammatic calculation}
In this section we present analytical formulas for the three density-dependent
strength functions $F_\tau(\rho)$, $F_{so}(\rho)$ and $F_J(\rho)$ as derived
(via the improved density-matrix expansion \cite{dmeimprov}) from 
$1\pi$-exchange, iterated $1\pi$-exchange, and irreducible $2\pi$-exchange 
diagrams with intermediate $\Delta$-isobar excitations, including 
Pauli-blocking corrections up to three-loop order. We give for each diagram
only the final result omitting all technical details related to extensive 
algebraic manipulations and solving elementary integrals. Further explanations 
about the organization and performance of our diagrammatic calculation can be 
found in section 3 of ref.\cite{efun}.   

\subsection{One pion exchange Fock diagram with two medium insertions}
The non-vanishing contributions from the $1\pi$-exchange Fock diagram shown in 
Fig.\,1, including the relativistic $1/M^2$-corrections, read: 
\begin{eqnarray}   F_\tau(\rho) &=& {3g_A^2 m_\pi \over (8 \pi f_\pi)^2 u^3}
  \bigg[ \Big(u^2 +{1\over 2}\Big) \ln(1+4u^2) -2u^2\bigg] \nonumber \\ && + 
{3g_A^2 m_\pi^3 \over (8 \pi f_\pi M)^2} \bigg[ 4u -{4u^3 \over 3} -2\arctan 2u 
-u \ln(1+4u^2)\bigg] \,, \end{eqnarray}
\begin{equation} F_J(\rho) =  {9g_A^2 \over (32 m_\pi f_\pi)^2 u^6} \Big[8u^4-4u^2 
+\ln(1+4u^2) \Big] \,,\end{equation}
where we have introduced the convenient dimensionless variable $u=k_f/m_\pi$.
\subsection{Iterated one-pion exchange Hartree diagram with two medium insertions} 
The two-body contributions from the iterated $1\pi$-exchange Hartree diagram
in Fig.\,1 read:  
\begin{equation} F_\tau(\rho) =  {g_A^4 M m_\pi^2 \over (8\pi)^3f_\pi^4}\bigg\{ 
{7+30u^2 \over 2u^3}\ln(1+4u^2)-{14 \over u} - 16\arctan 2u \bigg\} \,,
\end{equation}
\begin{equation} F_{so}(\rho) =  {3g_A^4 M \over \pi m_\pi (4f_\pi u)^4}\bigg\{ 
4u \arctan 2u -3u^2 -{5\over 4}\ln(1+4u^2)\bigg\} \,. \end{equation}
The expression for $F_{so}(\rho)$ in eq.(13) gives (part of) the ''wrong-sign'' 
spin-orbit interaction induced by the pion-exchange tensor force in second
order. It is completed by the Fock (exchange) contribution and the respective
Pauli-blocking corrections (see eqs.(15,18,22,25)).  
\subsection{Iterated one-pion exchange Fock diagram with two medium insertions}
We find the following contributions from the right diagram in Fig.\,1 with two
medium insertions on non-neighboring nucleon propagators:
\begin{eqnarray}  F_\tau(\rho) &=& {3g_A^4 M m_\pi^2 \over (4f_\pi)^4 (\pi u)^3}
\int_0^u \!\!dx \, {2x^2-u^2 \over 1+2x^2}  \nonumber \\ && \times
\Big[(1+8x^2+8x^4) \arctan x -(1+4x^2) \arctan 2x \Big] \,, \end{eqnarray}
\begin{eqnarray}  F_{so}(\rho) &=& {3g_A^4 M \over 2\pi m_\pi (4f_\pi u)^4}\bigg\{ 
u^2+ \int_0^u \!\!dx \, {1\over 1+2x^2}  \nonumber \\ && \times
\Big[4x^2(1+x^2) \arctan x -(1+4x^2) \arctan 2x \Big] \bigg\}\,, \end{eqnarray}
\begin{eqnarray}  F_J(\rho) &=& {9g_A^4 M \over \pi m_\pi (8f_\pi u)^4}\bigg\{ 
{2\over u^2} \int_0^u \!\!dx \, {1\over 1+2x^2} \Big[(u^2-x^2)(1+4x^2) 
 \arctan 2x \nonumber \\ && +2(u^2-2x^2+2u^2x^2-6x^4+2u^2x^4-6x^6)\arctan x 
\Big]-u^2 \bigg\}\,. \end{eqnarray}

\subsection{Iterated one-pion exchange Hartree diagram with three medium  insertions}
In our way of organizing the many-body calculation, the Pauli-blocking 
corrections are represented by diagrams with three medium insertions. The
corresponding contributions from the iterated $1\pi$-exchange Hartree diagram 
read: 
\begin{eqnarray}  F_\tau(\rho) &=& {3g_A^4 M m_\pi^2 \over (4\pi f_\pi)^4}\Bigg\{ 
2u^2-\ln(1+4u^2) +{2u^2 \over 1+4u^2} +2 \int_0^1 \!\!dy \,y^2\ln{1+y\over  1-y}
 \nonumber \\ && \times  \bigg[{8u^4 y^2 \over (1+4u^2y^2)^2}(6u^2y^2+y^2-2u^2) 
-4u^2y^2 +\ln(1+4u^2y^2)\bigg]  \nonumber \\ && +\int_0^u \!\! dx\,{x^2\over u^3}
\int_{-1}^1 \!\! dy\, \bigg[{2u x y\over u^2-x^2y^2}+\ln{u+x y \over u-x y}
\bigg] \bigg[ 2\ln(1+s^2)  \nonumber \\ &&-{2s^2+s^4 \over 1+s^2} +{(u^2-x^2y^2) 
s^4 \over u^2(1+s^2)^3} \Big((5+s^2)s'^2 +(s+s^3)(s''-2s')\Big) \bigg] \Bigg\}
\,, \end{eqnarray}
with the auxiliary function $s=xy +\sqrt{u^2-x^2+ x^2y^2}$, and its partial 
derivatives $s'=u\partial s/\partial u$ and $s''=u^2 \partial^2 s/\partial u^2$.
\begin{eqnarray}  F_{so}(\rho) &=& {3g_A^4 M \over \pi^2 m_\pi (4f_\pi)^4}
\int_0^u \!\! dx\,{x^2\over u^6} \int_{-1}^1 \!\! dy\,\Bigg\{\bigg[ 4x y 
\ln{u+x y \over u-x y} +{u(5x^2y^2-3u^2) \over  u^2-x^2y^2}\bigg]\nonumber \\ &&
\times \bigg[5s+{s\over (1+s^2)^2} -6 \arctan s\bigg] -{u s^5 (u^2+x^2y^2) 
\over (1+s^2)^2 (u^2-x^2y^2)}\nonumber \\ && +{2s^4 s'(s-2x y)\over (1+s^2)^2} 
\ln{u+x y \over u-x  y} \Bigg\}\,, \end{eqnarray}   
\begin{equation} F_J(\rho)={9g_A^4 M u^3\over 16 \pi^2 m_\pi f_\pi^4}\int_0^1\!\!dy 
\,{y^6 \over(1+4u^2y^2)^2} \bigg[ 2y+(1-y^2) \ln{1+y\over 1-y}\bigg] \,.
\end{equation}
\subsection{Iterated one-pion exchange Fock diagram with three medium insertions}
The evaluation of this diagram is most tedious. It is advisable to split the
contributions to the strength functions $F_\tau(\rho)$, $F_{so}(\rho)$ and 
$F_J(\rho)$ into ''factorizable'' and ''non-factorizable'' parts. These two 
pieces are distinguished by the feature of whether the nucleon propagator in
the denominator can be canceled or not by terms from the product of 
$\pi N$-interaction vertices in the numerator. We find the following 
''factorizable'' contributions:
\begin{eqnarray}F_\tau(\rho) &=&  {3g_A^4 M m_\pi^2 \over (4\pi f_\pi u)^4}\Bigg\{ 
{1\over 8}(1+4u^2+2u^4)\ln(1+4u^2)-{1+6u^2+8u^4 \over 64u^2}\ln^2(1+4u^2)
\nonumber \\ && -{u^4\over 2} -{u^2 \over 4}+{u\over 2}\int_0^u \!\!dx\, \Big[ 
u(1+u^2+x^2)-\Big(1+(u+x)^2\Big)\Big(1+(u-x)^2\Big)L\Big]\nonumber \\ &&
\times  \bigg[(1-u^2-x^2) L +u -{u \over 1+(u+x)^2} -{u \over 1+(u-x)^2}\bigg] 
\Bigg\}\,,\end{eqnarray} 
 with the auxiliary function:
\begin{equation} L(x,u)= {1\over 4x} \ln{1+(u+x)^2\over 1+(u-x)^2} \,.
\end{equation} 
\begin{eqnarray} F_{so}(\rho) &=&  {3g_A^4 M \over \pi^2 m_\pi (8f_\pi u)^4}\Bigg\{ 
4 \big[\ln(1+4u^2)-7u^2\big]\arctan 2u +28u^3+8u +{3\over u}\nonumber \\ &&-  
{3+14u^2+10u^4 \over 2u^3}\ln(1+4u^2)+{3+20u^2+16u^4 \over 16u^5}\ln^2(1+4u^2)
\nonumber \\ && +4\int_0^u \!\!dx\, \bigg\{L^2\Big[3x^{-2}(1+u^2)^3+3+2u^2-u^4-
(3+7u^2)x^2+5x^4\Big] \nonumber \\ && -6 u x^{-2}(1+u^2)^2 L +3u^2x^{-2}(1+u^2)
\bigg\} \Bigg\} \,,\end{eqnarray} 
\begin{eqnarray}  F_J(\rho) &=&  {9g_A^4 M \over \pi^2 m_\pi (8f_\pi u)^4}\Bigg\{ 
7\arctan 2u +4u^3 -{1\over u} -{1+4u^2\over 16u^5}\ln^2(1+4u^2) \nonumber \\ && 
-{37 u \over 4} +{8-7u^2-12u^4 \over 16u^3}\ln(1+4u^2) +\int_0^u \!\!dx\,  
\bigg\{{L^2 \over u^2}\bigg[ {3\over 2x^2}(1+u^2)^4 \nonumber \\ && +2(1-u^4)
(1+u^2) +(5+2u^2+5u^4)x^2 -(6+10u^2)x^4 +{11 x^6 \over 2}\bigg] \nonumber \\ &&
+{L\over u}\bigg[3u^4+2u^2-1-{3\over x^2}(1+u^2)^3\bigg] +{3\over 2x^2}(1+u^2)^2 
\bigg\} \Bigg\} \,.\end{eqnarray} 
The ''non-factorizable'' contributions (stemming from nine-dimensional principal
value integrals over the product of three Fermi-spheres of radius $k_f$) read
on the other hand: 
\begin{eqnarray} F_\tau(\rho) &=& {3g_A^4 M m_\pi^2 \over (4\pi f_\pi)^4} \int_{-1}^1
\!\!dy \int_{-1}^1 \!\!dz\, {yz \,\theta(y^2+z^2-1) \over |yz|\sqrt{y^2+z^2-1}}
\bigg\{ {4u^2z^2(2z^2-1) \over 1+4u^2z^2} \nonumber \\ &&
 \times \Big[ \ln(1+4u^2y^2)-4u^2y^2 \Big]\theta(y)\theta(z) + \int_0^u \!\!dx
\,{x^2 s^2 \over 2u^5(1+s^2)^2}\nonumber \\ && \times\Big[t^2- \ln(1+t^2)\Big] 
\Big[(s+s^3)(2s'-s'')-(3+s^2)s'^2\Big]  \bigg\}\,, \end{eqnarray}
\begin{eqnarray} F_{so}(\rho) &=& {3g_A^4 M \over \pi^2 m_\pi(4f_\pi)^4} \int_{-1}^1
\!\!dy \int_{-1}^1 \!\!dz\, {yz \,\theta(y^2+z^2-1) \over  |yz|\sqrt{y^2+z^2-1}}
\bigg\{{8 y^2 z\, \theta(y)\theta(z)\over 1+4u^2y^2}\nonumber \\ && \times  
\Big[\arctan(2uz )-2uz\Big]+ \int_0^u \!\! dx\,{x^2 s^2s't^2t' \over  2u^8(1+s^2)
(1+t^2)}(t x y -s x z-s t) \bigg\}\,, \end{eqnarray}
\begin{eqnarray} F_J(\rho) &=& {9g_A^4 M \over \pi^2 m_\pi(4f_\pi)^4} \int_{-1}^1
\!\!dy \int_{-1}^1 \!\!dz\, {yz \,\theta(y^2+z^2-1) \over  |yz|\sqrt{y^2+z^2-1}}
\bigg\{{2 y^4\,\theta(y)\theta(z)\over u(1+4u^2y^2)}\nonumber \\ && \times  
\Big[\ln(1+4u^2z^2)-4u^2z^2\Big]+ \int_0^u \!\! dx\,{x^4 s^3s't^3  t'(1-y^2-z^2) 
\over  4u^{10}(1+s^2)(1+t^2)}\bigg\}\,, \end{eqnarray}
with the auxiliary function $t=xz +\sqrt{u^2-x^2+x^2z^2}$ and its partial
derivative $t'=u \partial t/\partial u$. For the numerical evaluation of the
$dy\,dz$-double integrals in eqs.(24,25,26) it is advantageous to first
antisymmetrize the integrands in $y$ and $z$ and then to substitute $z=
\sqrt{y^2\zeta^2+1-y^2}$. This way the integration region becomes equal to the
unit-square $0<y,\zeta<1$. 
\subsection{Irreducible two-pion exchange}
At next order in the small momentum expansion comes the irreducible
$2\pi$-exchange including (also) intermediate $\Delta$-isobar excitations. We 
employ a (subtracted) spectral-function representation of the 
$\pi N\Delta$-loops and find the following (two-body) contributions:   
\begin{eqnarray} F_\tau(\rho) &=& {1\over 8\pi^3} \int_{2m_\pi}^\infty\!\! d\mu
\,{\rm Im}(V_C+3W_C+2\mu^2 V_T+6\mu^2 W_T) \nonumber \\ && \times \bigg[ {2\mu
  \over k_f} +{8k_f^3 \over 3 \mu^3} -{\mu \over 2k_f^3}(\mu^2+2k_f^2)
\ln\bigg(1+{4k_f^2 \over \mu^2}\bigg) \bigg] \,, \end{eqnarray}   
\begin{eqnarray} F_J(\rho) &=& {3\over 16\pi} \int_{2m_\pi}^\infty\!\! d\mu\, \Bigg\{
{\rm Im}(V_C+3W_C)\bigg[{\mu \over 4k_f^6}(\mu^2+2k_f^2)\ln\bigg(1+{4k_f^2 
\over \mu^2}\bigg)-{\mu  \over k_f^4}-{4\over 3 \mu^3}\bigg] \nonumber \\ && +
{\rm Im}(V_T+3W_T)\bigg[ {\mu  \over k_f^2} -{4 \over 3\mu}+{\mu^3 \over 2k_f^4} 
-{\mu^3 \over 8k_f^6}(\mu^2+4k_f^2)\ln\bigg(1+{4k_f^2 \over \mu^2}\bigg)
\bigg] \Bigg\}\,. \end{eqnarray}   
The imaginary parts Im$V_C$, Im$W_C$, Im$V_T$ and Im$W_T$ of the isoscalar and
isovector central and tensor NN-amplitudes due to $2\pi$-exchange with single
and double $\Delta$-excitation can be found in section 3 of 
ref.\cite{spectral}. The additional contributions from the irreducible 
$2\pi$-exchange with only nucleon intermediate states are accounted for by
inserting into eqs.(27,28) the imaginary parts:
\begin{equation} {\rm Im}W_C= {\sqrt{\mu^2-4m_\pi^2} \over 3\pi 
\mu (4f_\pi)^4} \bigg[ 4m_\pi^2(1+4g_A^2-5g_A^4) +\mu^2(23g_A^4-10g_A^2-1) + 
{48 g_A^4 m_\pi^4 \over \mu^2-4m_\pi^2} \bigg] \,, \end{equation}
\begin{equation} {\rm Im}V_T= - {6 g_A^4 \sqrt{\mu^2-4m_\pi^2} \over 
\pi  \mu (4f_\pi)^4}\,. \end{equation}
At leading order the irreducible $2\pi$-exchange generates no spin-orbit
NN-interaction. It emerges first as a relativistic $1/M$-correction. In order to
see the size of such relativistic effects we have evaluated the energy 
density functional with a two-body interaction composed of the (isoscalar and
isovector) spin-orbit NN-amplitudes $V_{\rm SO}$ and $W_{\rm SO}$ written in
eqs.(22,23) of ref.\cite{nnpap}. We find with it the following contribution to 
the spin-orbit coupling strength:       
\begin{eqnarray} F_{so}(\rho) &=&{g_A^2 m_\pi\over\pi M(4f_\pi)^4}\bigg\{\bigg(g_A^2 
-{4\over 5}\bigg) \bigg[ {1\over u^4}\ln(1+u^2) -{1\over u^2}\bigg] \nonumber \\
&& +{18 \over 5} - {3g_A^2 \over 2}+\bigg( {2g_A^2 -4 \over u}-{12u \over 5}
\bigg) \arctan u \bigg\}\,, \end{eqnarray}   
which has been subtracted at $\rho=0$ in order to eliminate (regularization
dependent) short-distance components. As a consequence of that subtraction only
the Fock terms are included in the expressions in eqs.(27,28,31).
\subsection{Three-body diagrams with $\Delta$-excitation}
The Pauli-blocking correction to the $2\pi$-exchange with single 
$\Delta$-excitation is equivalent to the contribution of a (genuine)
three-nucleon force. In fact, one is dealing here with the same three-nucleon 
interaction as originally introduced by Fujita and Miyazawa \cite{fujita}.
Moreover, it has been shown in ref.\cite{achim} that the inclusion of this
long-range $2\pi$-exchange three-nucleon interaction is essential in order to
reproduce the empirical saturation point of nuclear matter when using the
low-momentum NN-potential $V_{\rm low-k}$ in Hartree-Fock calculations. It is
therefore equally interesting to see its effects on the nuclear energy 
functional ${\cal E}[\rho, \tau,\vec J\,]$.
 
\begin{figure}
\begin{center}
\includegraphics[scale=1.0,clip]{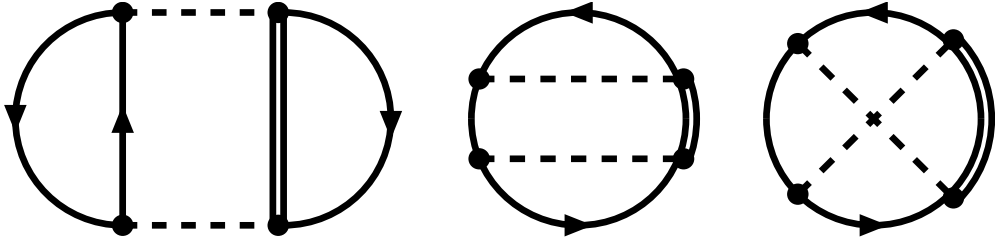}
\end{center}
\vspace{-0.2cm}
\caption{Hartree and Fock three-body diagrams related to $2\pi$-exchange with 
virtual $\Delta$-isobar excitation. Their isospin factors for 
isospin-symmetric nuclear systems are 8, 0, and 8, respectively.}
\end{figure}

The pertinent Hartree and Fock three-body diagrams related to $2\pi$-exchange
with virtual $\Delta$-excitation are shown in Fig.\,2. The central diagram
with parallel pion-lines vanishes for isospin-symmetric nuclear systems. 
Returning to the medium insertion written in eq.(5) we find from the left
three-body Hartree diagram in Fig.\,2 the following contributions: 
\begin{equation}   F_\tau(\rho) ={ g_A^4 m_\pi^4 \over \Delta(2 \pi f_\pi)^4}
\bigg\{\Big(u^2 +{3\over 4}\Big) \ln(1+4u^2) -{u^2(3+10u^2)\over 1+4u^2 }\bigg\} 
\,, \end{equation}
\begin{equation}   F_{so}(\rho) ={ 3g_A^4 m_\pi \over \pi^2\Delta(4f_\pi)^4}
\bigg\{ {12\over u}+8u -{3+8u^2 \over u^3} \ln(1+4u^2) \bigg\}\,,\end{equation}
\begin{equation}   F_J(\rho) ={ 3g_A^4 m_\pi \over \pi^2\Delta(4f_\pi)^4}
\bigg\{{3\over 2u^3} \ln(1+4u^2)+4u -{6\over u}+{8u \over 1+4u^2} \bigg\}
\,,\end{equation}
with $\Delta = 293\,$MeV the delta-nucleon mass splitting. We have used the 
value $3/\sqrt{2}$ for the ratio between the $\pi N\Delta$- and 
$\pi NN$-coupling constants. Note that the expression for $F_{so}(\rho)$ in 
eq.(33) gives the (dominant part of the) three-body spin-orbit coupling 
strength suggested originally by Fujita and Miyazawa \cite{fujita}. The 
three-body effects on the energy density functional are completed by the 
contributions from the right Fock diagram in Fig.\,2 which read:
\begin{eqnarray}F_\tau(\rho) &=& { g_A^4 m_\pi^4\over \Delta(4 \pi f_\pi)^4}\Bigg\{ 
{2\over u^3}\Big[4u^2-(1+2u^2) \ln(1+4u^2)\Big] \arctan 2u -{2u^4 \over 3}
\nonumber \\ && +{31 u^2 \over 6}- {31 \over 4}+{5\over 8u^2} +{3+22u^2+176u^4
+288 u^6 \over 256 u^8 }\ln^2(1+4u^2)\nonumber \\ && + {3\over 16u^4}+ {1\over
  96 u^6}( 248u^6-224u^8-60u^4-48u^2-9) \ln(1+4u^2) \nonumber \\ && 
+{1\over u^3} \int_0^u\!\!dx\bigg\{G_S\bigg[{2u(u+x)\over 1+(u+x)^2}+{2u(x-u)
\over 1+(u-x)^2} - 4 x L\bigg]\nonumber \\ && +G_T \bigg[{3u \over 4x}(3u^2-1)
-{3u x \over 4}+{u(u+x)\over 1+(u+x)^2}+{u(x-u) \over 1+(u-x)^2} \nonumber \\
&& +{L \over 4x} (3x^4+6u^2x^2-2x^2-9u^4-6u^2+3) \bigg] \bigg\}\Bigg\} \,, 
\end{eqnarray}
with the auxiliary functions: 
\begin{eqnarray} G_S(x,u) &=& {4ux \over 3}( 2u^2-3) +4x\Big[
\arctan(u+x)+\arctan(u-x)\Big] \nonumber \\ && + (x^2-u^2-1) \ln{1+(u+x)^2
\over  1+(u-x)^2} \,,\end{eqnarray}
\begin{eqnarray} G_T(x,u) &=& {ux\over 6}(8u^2+3x^2)-{u\over
2x} (1+u^2)^2  \nonumber \\ && + {1\over 8} \bigg[ {(1+u^2)^3 \over x^2} -x^4 
+(1-3u^2)(1+u^2-x^2)\bigg] \ln{1+(u+x)^2\over  1+(u-x)^2} \,.\end{eqnarray}
\begin{eqnarray}F_{so}(\rho) &=& { g_A^4 m_\pi\over \pi^2\Delta(8 f_\pi)^4}\Bigg\{ 
{64u \over 3}-{4\over u}-{7\over u^3}-{12\over u^5}-{15\over 4u^7} \nonumber
\\ && + \bigg({15\over 8u^9}+ {39\over 4u^7}+ {13\over u^5} + {6\over u^3} - 
{8\over u} \bigg)\ln(1+4u^2) \nonumber \\ && -{3\over 64 u^{11}}
(64u^6+80u^4+36u^2+5) \ln^2(1+4u^2) \Bigg\}   \,,\end{eqnarray}
\begin{eqnarray}F_J(\rho) &=& { g_A^4 m_\pi\over \pi^2 \Delta(8 f_\pi u)^4}\Bigg\{ 
24\Big[4-8u^2-{1\over u^2} \ln(1+4u^2)\Big] \arctan 2u - 144u^5  +{3\over u}
\nonumber \\ && +272u^3-99u +{9\over 4u^3}+\bigg(28u-{15\over 4u^3}-{9\over
8 u^5}\bigg) \ln(1+4u^2) + {3\over 64 u^7}\nonumber \\ &&\times (3+16u^2 +144u^4)
\ln^2(1+4u^2) +{3\over 4}(58u^4+31u^2-63) \arctan 2u \nonumber \\ && +{663 u^3
\over 16}+{495u \over 16}-{656 u^5\over 5} +{9 \over 64u}(29-229u^2+52u^4)
\ln(1+4u^2) \nonumber \\ && +\int_0^u\!\!dx\,\bigg\{{9L^2 \over 8u^2}\bigg[
{6\over x^2}(1+u^2)^4(3u^2-1)-{5\over x^4}(1+u^2)^6+(1+u^2)^2\nonumber \\ &&
\times (50u^2-39-55u^4) + 4x^2(35u^6+5u^4-39u^2-9) -33x^8 \nonumber \\ && +13x^4
(1+2u^2-15u^4) +2x^6(65u^2-11)\bigg]+{3L \over 4u}\bigg[{15\over  x^4}(1+u^2)^5
 \nonumber \\ && +{1\over x^2}(1+u^2)^3(3-49u^2) +6(25u^6+5u^4-u^2+19) \bigg]
\nonumber \\ && -{45\over  8x^4}(1+u^2)^4 +{3\over  2x^2}(1+u^2)^2(3+11u^2) 
\bigg\}\Bigg\}   \,.\end{eqnarray}
In the appendix we present the three-body spin-orbit coupling strength 
$F_{so}(\rho)$ for an alternative description of the $2\pi$-exchange
three-nucleon interaction (using $\pi\pi NN$-contact vertices instead of
propagating $\Delta$-isobars). A good check of all formulas collected in 
this section is provided by their Taylor series expansion in $k_f$. Despite the
superficial opposite appearance the leading term in the $k_f$-expansion is 
always a non-negative power of $k_f$ (which is higher for three-body 
contributions than for two-body contributions). 

\section{Results and discussion}
In this section we present and discuss our numerical results obtained by
summing the series of contributions given in section 3. The physical input
parameters are: $g_A=1.3$ (nucleon axial vector coupling constant), $f_\pi = 
92.4\,$MeV (pion decay constant), $m_\pi = 135\,$MeV (neutral pion mass) and 
$M=939\,$MeV (nucleon mass). We recall that with these physical parameters and 
a few adjustable short-distance couplings the nuclear matter equation of state 
$\bar E(\rho)$ and many other nuclear matter properties \cite{deltamat} can be 
well described by the chiral pion-nucleon dynamics treated to three-loop order. 

Returning to the energy density functional ${\cal E}[\rho,\tau,\vec J\,]$ in
eq.(6) one observes that the expression multiplying the kinetic energy density
$\tau(\vec r\,)$ has the meaning of a reciprocal density-dependent effective
nucleon mass:    
\begin{equation} M^*(\rho) = M\bigg[1-{k_f^2 \over 2M^2} +2M  F_\tau(\rho)
\bigg]^{-1}\,. \end{equation}
It is identical to the so-called ''Landau'' mass introduced in Fermi-liquid 
theory, since it derives in the same way from the slope of the single-particle 
potential $U(p,k_f)$ at the Fermi-surface $p=k_f$. This consistency of 
effective nucleon masses follows from the improved density-matrix expansion 
of  Gebremariam, Duguet and Bogner \cite{dmeimprov}, but it did not hold for 
the original density-matrix expansion of Negele and Vautherin 
\cite{negele,efun}. 
\begin{figure}
\begin{center}
\includegraphics[scale=0.55,clip]{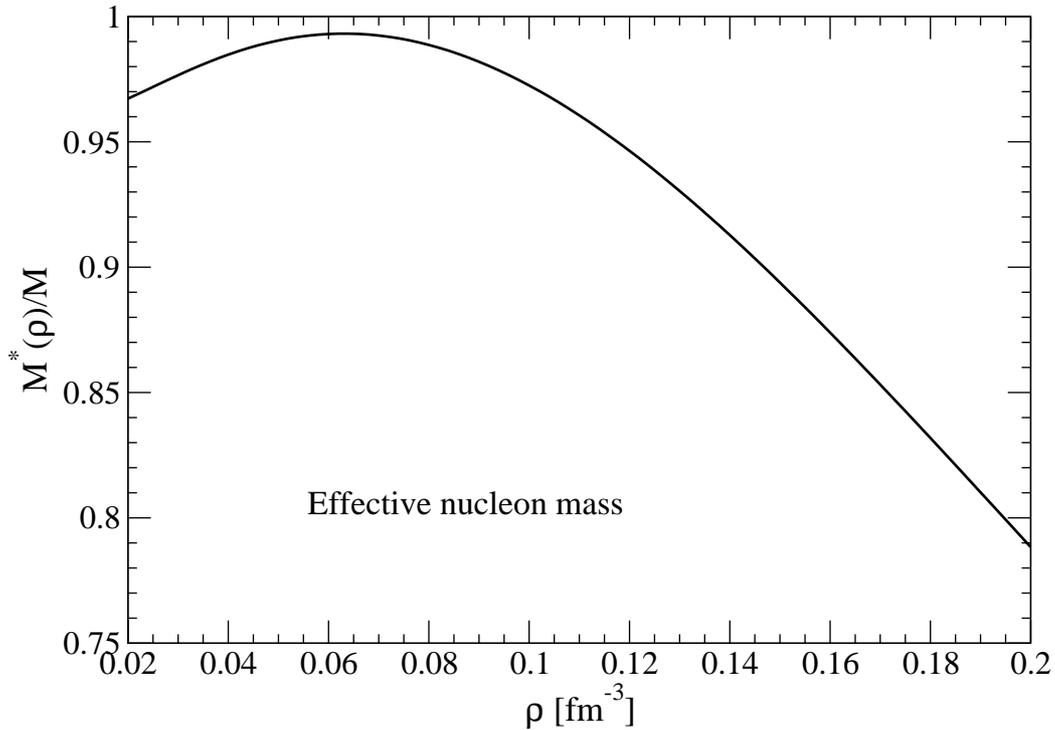}
\end{center}
\vspace{-0.2cm}
\caption{The effective nucleon mass $M^*(\rho)$ divided by the free nucleon mass
$M$ as a function of the nuclear density $\rho$.}
\end{figure}

Fig.\,3 shows the ratio of effective to free nucleon mass $M^*(\rho)/M$ as a
function of the nuclear density $\rho = 2k_f^3/3\pi^2$. One observes a reduced
effective nucleon mass which reaches the value $M^*(\rho_0)  =0.874M$ at
nuclear matter saturation density $\rho_0 =0.16\,$fm$^{-3}$. This is compatible 
with the range $0.7 <M^*(\rho_0)/M<1$ spanned by phenomenological Skyrme
forces \cite{sk3,skmstar,sly,pearson}. Somewhat unusual is the non-monotonic 
progression of the curve in Fig.\,3. It reveals that a sufficiently high 
density $(0.4\rho_0)$ has to be reached until the subleading $\pi N
\Delta$-dynamics can revert the tendency of the iterated $1\pi$-exchange to 
increase the effective nucleon mass. The same feature has also been observed 
for the $p$-wave Landau parameter $f_1(k_f)$ in ref.\cite{quasi} (see Fig.\,3 
therein), a quantity which is intimately related to the effective nucleon mass 
$M^*(\rho)$.
\begin{figure}
\begin{center}
\includegraphics[scale=0.55,clip]{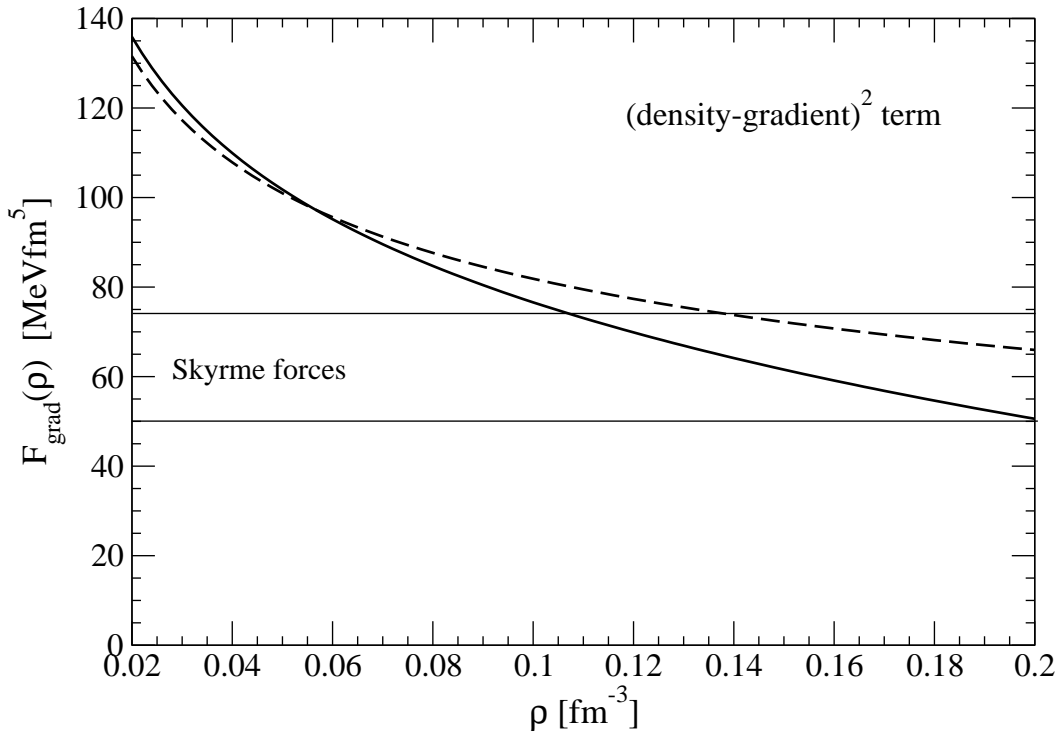}
\end{center}
\vspace{-0.2cm}
\caption{The strength function $F_\nabla(\rho)$ of the surface-term 
$(\vec \nabla \rho)^2$ in the nuclear energy density functional versus 
the nuclear density $\rho$. The dashed line corresponds to the
truncation to $1\pi$- and iterated $1\pi$-exchange. The full line includes
also $2\pi$-exchange and associated three-body contributions.}
\end{figure}

Next, we show in Fig.\,4 the strength function $F_\nabla(\rho)$ belonging to
the $(\vec \nabla \rho)^2$ surface-term. The dashed line corresponds to the
truncation to $1\pi$- and iterated $1\pi$-exchange, whereas the full line 
includes in addition the $2\pi$-exchange and the associated three-body 
contributions. Taking the band spanned by phenomenological Skyrme forces 
\cite{sk3,skmstar,sly,pearson} as a benchmark one may conclude that the 
subleading $2\pi$-exchange dynamics leads to some improvement. The improved 
density-matrix expansion \cite{dmeimprov} has furthermore flattened and
shifted downward the curve for $F_\nabla(\rho)$ in comparison to our previous
calculation (see Fig.\,8 in ref.\cite{deltamat}) based on the Negele-Vautherin
density-matrix expansion. We also note that in the relevant density region
$\rho_0/2 < \rho < \rho_0$ the main contribution to the strength function 
$F_\nabla(\rho)$ comes from the component $F_d(\rho)$ (see eq.(8)) which is
insensitive to the density-matrix expansion beyond its fixed nuclear matter 
part $\theta(k_f-|\vec p\,|)$.           
\begin{figure}
\begin{center}
\includegraphics[scale=0.55,clip]{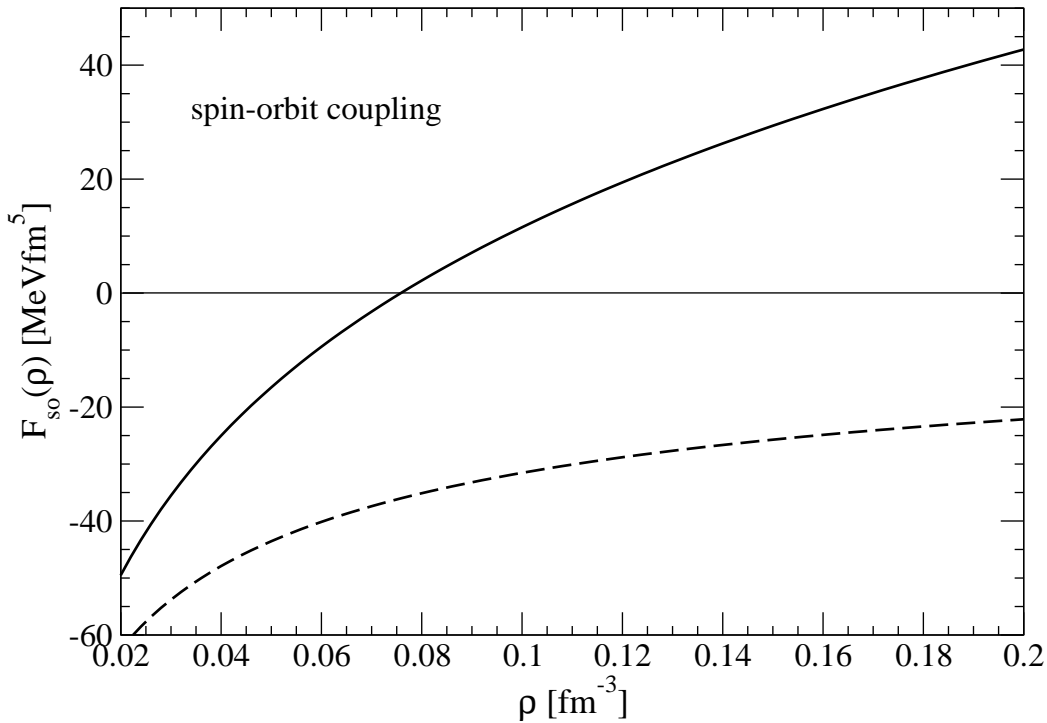}
\end{center}
\vspace{-0.2cm}
\caption{The strength function $F_{so}(\rho)$ of the spin-orbit
coupling term $\vec \nabla \rho \cdot \vec J$ in the nuclear energy density 
functional versus the nuclear density $\rho$. Dashed line: $1\pi$- and 
iterated $1\pi$-exchange only. Full line: $2\pi$-exchange and three-body 
contributions added.}
\end{figure}

Of particular interest is the strength $F_{so}(\rho)$ of the spin-orbit
coupling provided by the explicit pion-exchange dynamics.  The dashed curve in
Fig.\,5 shows the ''wrong-sign'' spin-orbit coupling strength arising from
iterated $1\pi$-exchange (i.e. the pion-exchange tensor force in second order). 
Its value at half nuclear matter density $\rho_0/2 = 0.08\,$fm$^{-3}$ decomposes 
as $((-124.4+47.7)+(76.5-35.0))\,$MeVfm$^5$ $=-35.1\,$MeVfm$^5$ into Hartree and
Fock pieces supplemented by the respective Pauli-blocking corrections. This
net negative result amounts to about $-40\%$ of the empirical spin-orbit
coupling strength $F_{so}^{(\rm emp)} \simeq 90\,$MeVfm$^5$. In comparison to our 
previous calculation \cite{efun} based on the 
Negele-Vautherin density-matrix expansion which gave at $\rho_0/2$ the value 
$-58.1\,$MeVfm$^5$ the ''wrong-sign'' spin-orbit coupling strength has 
substantially decreased in magnitude. The full line in Fig.\,5 shows the 
spin-orbit coupling strength after including the subleading $2\pi$-exchange, 
in particular the three-body contributions eqs.(33,38). One finds now a 
pronounced cancellation in the density region around $\rho_0/2=0.08\,$fm$^{-3}$,
where the spin-orbit interaction in nuclei gains actually most of its weight. 
Such an almost complete cancellation leaves then all room for the 
short-distance NN-dynamics (not treated explicitly in this work) to account 
for the strong spin-orbit coupling in nuclei. In fact, it has been shown in 
ref.\cite{short} that the empirical value $F_{so}^{(\rm  emp)} \simeq 90 
\,$MeVfm$^5$ of the spin-orbit coupling strength in nuclei is in perfect 
agreement with the one extracted from  realistic nucleon-nucleon potentials. 
The intimate connection between the strong Lorentz scalar and vector 
mean-fields and the (short-range) spin-orbit part of the NN-potential has been 
elucidated in ref.\cite{tueb} via (relativistic) Dirac-Brueckner calculations 
of the in-medium nucleon self-energy.  

Moreover, we note that the spin-orbit coupling strength generated by the 
irreducible $2\pi$-exchange as a relativistic $1/M$-correction (see eq.(31) in 
section 3.6) contributes little to the cancellation between "wrong-sign" and 
"correct-sign" parts shown in Fig.\,5. At $\rho_0/2 = 0.08\,$fm$^{-3}$ this 
piece amounts to just about  $-4.0\,$MeVfm$^5$. Furthermore, we have convinced 
ourselves that the spin-orbit NN-amplitudes from $2\pi$-exchange with 
$\Delta$-excitation ($V_{SO}$ and $W_{SO}$ collected in the appendix of 
ref.\cite{spectral}) lead to an even smaller effect. These  NN-amplitudes make 
up a two-body contribution to $F_{so}(\rho)$ that scales again with $1/M$.

\begin{figure}
\begin{center}
\includegraphics[scale=0.55,clip]{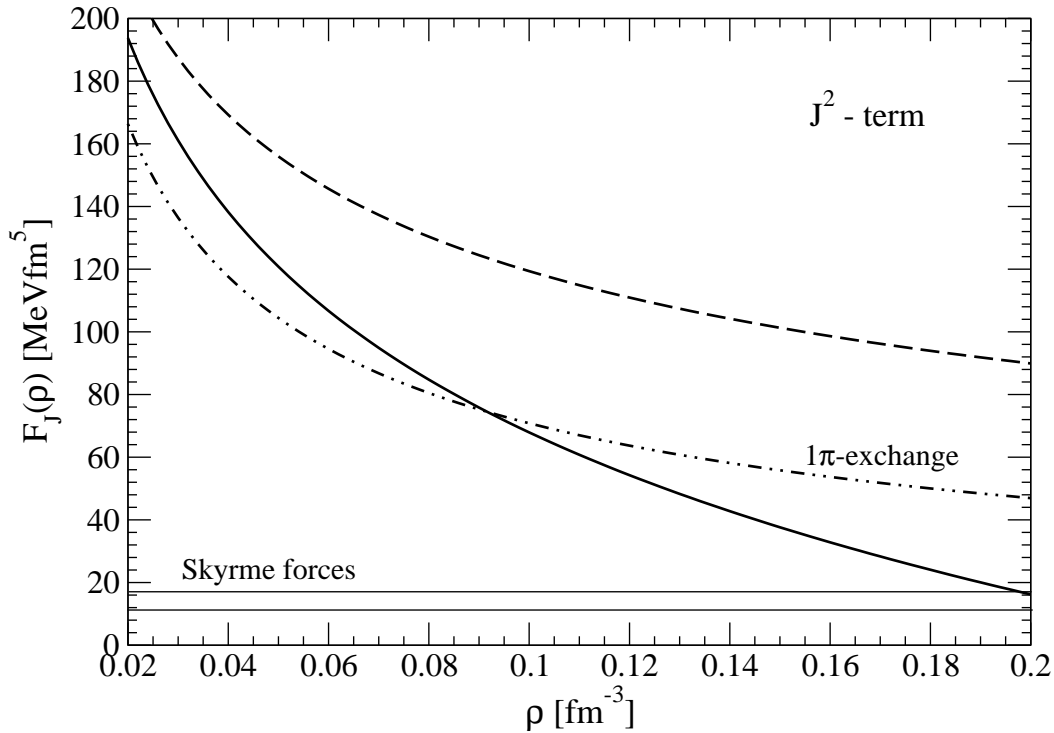}
\end{center}
\vspace{-0.2cm}
\caption{The strength function $F_J(\rho)$ accompanying the squared 
spin-orbit density $\vec J\,^2$ in the nuclear energy density functional 
versus the nuclear density $\rho$. Dashed line: $1\pi$- and 
iterated $1\pi$-exchange only. Full line: $2\pi$-exchange and three-body 
contributions added.}
\end{figure}

Finally, we show in Fig.\,6 the strength function $F_J(\rho)$ belonging the 
squared spin-orbit density $\vec J\,^2$ in the nuclear energy density 
functional as a function of the nuclear density $\rho$. One observes that the
inclusion of the subleading $2\pi$-exchange strongly reduces the values of 
$F_J(\rho)$. In comparison to the (narrow) band spanned by phenomenological 
Skyrme forces \cite{sk3,skmstar,sly,pearson} our prediction for the strength
function $F_J(\rho)$ is much larger in the whole density region  $0 < \rho <
\rho_0$. In addition, the density dependence of $F_J(\rho)$ comes out
markedly different, due to the long-range character of the pion-exchange 
interactions. For orientation, we reproduce by the dashed-dotted line
in Fig.\,6 the leading contribution from the $1\pi$-exchange Fock diagram (see 
eq.(11)). We also note that in comparison to the calculation based on the 
Negele-Vautherin density-matrix expansion the magnitude of the strength
function $F_J(\rho)$ has substantially increased (see Fig.\,5 in 
ref.\cite{efun}).   

Besides representing the non-local Fock contributions from tensor forces etc. 
in the energy density functional the $\vec J^{\,2}$-term leads to another 
interesting side effect. Namely, it gives rise to an extra spin-orbit 
single-particle mean-field $2F_J(\rho)\, \vec J$  in addition to the ''normal'' 
one, $F_{so}(\rho)\, \vec \nabla \rho$. It would be interesting to investigate
the role of this additional (nucleus-dependent) spin-orbit mean-field together
with the large values and the strong density dependence of $F_J(\rho)$ as 
predicted by in-medium chiral perturbation theory.

\section{Summary and concluding remarks}
In this work we have used the recently improved density-matrix expansion of
Gebremariam, Duguet and Bogner \cite{dmeimprov} to calculate the nuclear
energy density functional ${\cal E}[\rho,\tau,\vec J\,]$ relevant for $N=Z$ 
even-even nuclei in the framework of in-medium chiral perturbation theory.  
Our calculation treats systematically the effects from $1\pi$-exchange,
iterated $1\pi$-exchange, and irreducible $2\pi$-exchange with intermediate 
$\Delta$-isobar excitations, including Pauli-blocking corrections up to 
three-loop order. 

We find that the effective nucleon mass $M^*(\rho)$ 
entering the energy density functional becomes identical to the one of 
Fermi-liquid theory when employing the improved density-matrix expansion. 
The strength $F_\nabla(\rho)$ of the $(\vec\nabla \rho)^2$ surface-term as 
provided by the pion-exchange dynamics is in good agreement with that of 
phenomenological Skyrme forces in the density region $\rho_0/2 <\rho < \rho_0$. 

The spin-orbit coupling strength $F_{so}(\rho)$ receives contributions from 
iterated $1\pi$-exchange (of the ``wrong sign'') and from three-nucleon 
interactions mediated by $2\pi$-exchange with virtual $\Delta$-excitation (of 
the ``correct sign''). In the region around $\rho_0/2 \simeq 0.08\,$fm$^{-3}$ 
where the spin-orbit interaction in nuclei gains most of its weight these two 
components tend to cancel, thus leaving all room for the short-range spin-orbit 
interaction. The empirical value $F_{so}^{(\rm emp)} \simeq 90\,$MeVfm$^5$ of
the spin-orbit coupling strength in nuclei agrees perfectly with the one
extracted from the short-range spin-orbit component of realistic NN-potentials 
\cite{short}. This part of the NN-interaction drives at the same time the
strong Lorentz scalar and vector mean-fields on which the whole success of the 
relativistic Dirac phenomenology rests.     

The strength function $F_J(\rho)$ multiplying the squared spin-orbit density 
comes out much larger than from phenomenological Skyrme forces and it has a 
pronounced density dependence due to the long-range character of the 
pion-exchange interaction. The interplay between the two components of 
the total nuclear spin-orbit mean-field $2F_J(\rho)\,\vec J +F_{so}(\rho)\, \vec 
\nabla \rho$ should be further explored together with the large values and
strong density dependence of $F_J(\rho)$ as predicted by in-medium chiral 
perturbation theory.

In comparison to refs.\cite{efun,deltamat} where the density-matrix expansion
of Negele and Vautherin has been employed, we find an improved description of 
the nuclear energy density functional ${\cal E}[\rho,\tau,\vec J\,]$ on the 
basis of the improved density-matrix expansion \cite{dmeimprov}. In view of the 
fact that short-range contributions do not change (as exemplified here for 
the Skyrme force), a cancellation of the net two-pion exchange spin-orbit 
coupling strength around half nuclear matter density $\rho_0/2=0.08\,$fm$^{-3}$ 
is more satisfactory than having this cancellation around $\rho_0$ 
as discussed in ref.\cite{note}. In any case, the effective field theory 
formulation of nuclear forces provides short-range contributions to all four 
strength  functions $F_\tau(\rho)$, $F_\nabla(\rho)$, $F_{so}(\rho)$ and $F_J(\rho)$
and these can be fine-tuned in nuclear structure calculations.    
\section*{Appendix: Three-body spin-orbit coupling strength}
\begin{figure}
\begin{center}
\includegraphics[scale=0.55,clip]{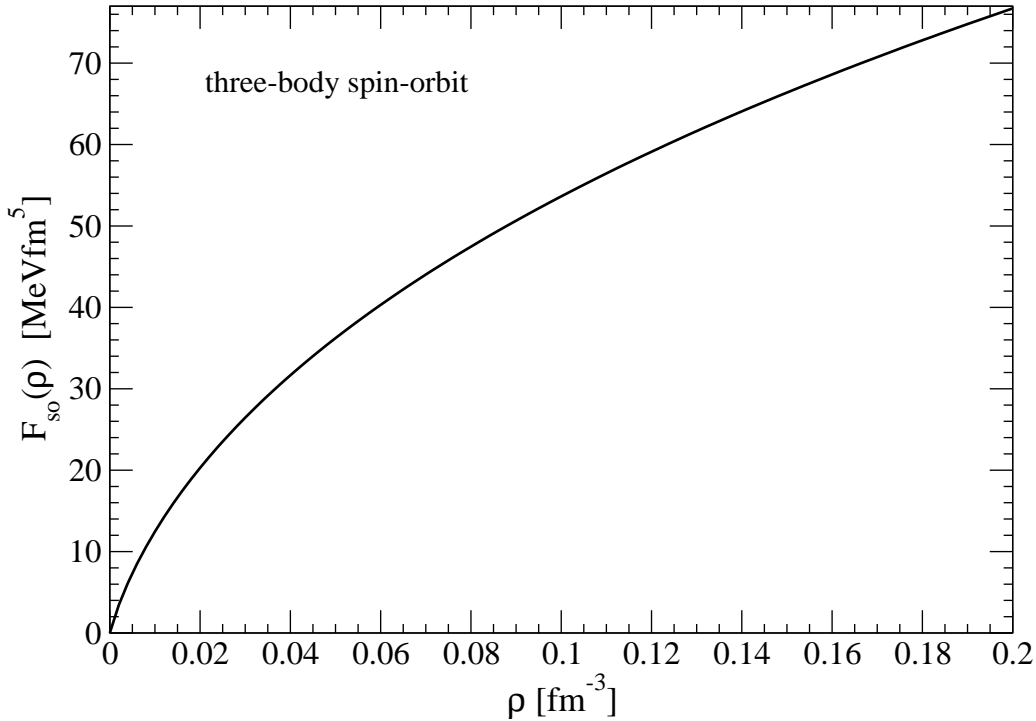}
\end{center}
\vspace{-0.2cm}
\caption{Three-body spin-orbit coupling strength $F_{so}(\rho)$ as a 
function of the nuclear density $\rho$.}
\end{figure}

In this appendix we present and discuss the result for the three-body 
spin-orbit coupling  strength $F_{so}(\rho)$ one obtains from an alternative 
description of the $2\pi$-exchange three-nucleon interaction. Instead of the 
sequential $\pi N \to \Delta \to \pi N$ transition with intermediate 
$\Delta$-isobar excitation one can employ the second order chiral $\pi\pi NN
$-contact vertex \cite{3bodyso}:
\begin{equation} {i\over f_\pi^2}\Big\{2\delta_{ab}(c_3 \vec q_a \cdot \vec  q_b 
-2c_1 m_\pi^2)+c_4 \, \epsilon_{abc} \tau_c \, \vec \sigma \cdot( \vec q_a \times
\vec q_b  ) \Big\}\,, \end{equation}
to built up the $2\pi$-exchange three-nucleon interaction. Here, $\vec q_{a,b}$ 
denote out-going pion momenta and we have already dropped the $c_2$ term 
proportional to the product of two pion energies. In the present application 
these (off-shell) pion energies are equal to differences of nucleon kinetic 
energies, thus producing a relativistic $1/M^2$-correction. The pertinent 
in-medium diagrams are those shown in Fig.\,2 with the $\Delta$-propagator 
shrunk to a point. Returning to the medium insertion written in eq.(5) we find 
from the corresponding three-body Hartree diagram (with two closed nucleon
rings) the following contribution to the spin-orbit coupling  strength:   
\begin{equation} F_{so}(\rho)  = {3g_A^2 m_\pi \over (8\pi)^2 f_\pi^4}
\bigg\{{2\over u}(4c_1-3c_3) -4c_3 u+\bigg[{4\over u}(c_3-c_1)+{3c_3-4c_1
    \over 2u^3}\bigg] \ln(1+4u^2) \bigg\}\,, \end{equation}
where $u = k_f/m_\pi$. It is completed by the contribution of the three-body 
Fock diagram (with a single closed nucleon ring) which reads:
\begin{eqnarray}F_{so}(\rho) &=& {g_A^2 m_\pi \over \pi^2 (4 f_\pi u)^4}\Bigg\{ 
3c_1\bigg[ 2u-2u^3+{3\over 2u} -{3+10u^2 \over 4u^3} \ln(1+4u^2) \nonumber \\
&& +{3+16u^2+16u^4 \over 32 u^5} \ln^2(1+4u^2)\bigg] + (c_3+c_4) \bigg[u^3-{16 
u^5\over 3} +{7u \over 4}\nonumber \\ && +{3 \over u}+{15\over 16u^3}+\bigg(
2u^3-{3u\over 2}-{13\over 4u}-{39 \over 16u^3}-{15\over  32u^5}\bigg)\ln(1+4u^2)
\nonumber \\ && +{3\over 256 u^7}(64u^6+80u^4+36u^2+5)  \ln^2(1+4u^2)\bigg]
\Bigg\}\,.\end{eqnarray}

Fig.\,7 shows the total three-body spin-orbit coupling strength $F_{so}(\rho)$ 
as a function of the nuclear density $\rho = 2k_f^3/3\pi^2$ for the choice of
low-energy constants: $c_1 = -0.81\,$GeV$^{-1}$,  $c_3 = -3.2\,$GeV$^{-1}$ and   
 $c_4 = 5.4\,$GeV$^{-1}$. Its value at half nuclear matter density, $F_{so}(
\rho_0/2) =47.5\,$MeVfm$^5$, is now about $14\%$ larger than the analogous 
$\Delta$-driven three-body effects presented in section 3.7. This small 
increase comes mainly from the fact that the low-energy constant $-c_3= 3.2\,
$GeV$^{-1}$ is not completely saturated by its dominant $\Delta$-resonance 
contribution $g_A^2/2\Delta = 2.9\,$GeV$^{-1}$. 

These considerations indicate that there is some uncertainty for the density 
at which the cancellation between ''wrong-sign'' and ''correct-sign'' 
spin-orbit coupling strength actually happens. Nevertheless, the general 
features of such a balance remain unchanged.


\begin{thebibliography}{99}
\bibitem{reinhard} M. Bender, P.H. Heenen and P.G. Reinhard, {\it Rev. Mod.
Phys.} {\bf  75} (2003) 121;\\ J.R. Stone  and P.G. Reinhard, {\it Prog. 
Part. Nucl. Phys.} {\bf  58} (2007) 587.\vs
\bibitem{skyrme} T.H.R. Skyrme, \textit{Nucl. Phys.} \textbf{9} (1959) 615.\vs
\bibitem{sk3} M. Beiner, H. Flocard, N. Van Giai and P. Quentin, \textit{Nucl. 
Phys.} \textbf{A238} (1975) 29.\vs
\bibitem{skmstar} J. Bartel, P. Quentin, M. Brack, C. Guet and H.B. Hakansson,
\textit{Nucl. Phys.} \textbf{A386} (1982) 79.\vs
\bibitem{sly} E. Chabanat, P. Bonche, P. Haensel, J. Meyer and R. Schaeffer,
\textit{Nucl. Phys.} \textbf{A627} (1997) 710; \textbf{A635} (1998) 231; 
and refs. therein.\vs
\bibitem{pearson} S. Goriely, M. Samyn, M. Bender and J.M. Pearson, {\it
Phys. Rev.}  {\bf C68} (2003) 054325;\\ M. Samyn,  S. Goriely, M. Bender and 
J.M. Pearson, {\it Phys. Rev.}  {\bf C70} (2004) 044309; \\ N. Chamel, 
S. Goriely and J.M. Pearson, {\it  Nucl. Phys.} {\bf A812} (2008) 72.\vs
\bibitem{walecka} B.D. Serot and J.D. Walecka, {\it Int. J. Mod. Phys.} {\bf 
E6} (1997) 515; and refs. therein.\vs 
\bibitem{ringreview} P. Ring, Lecture Notes in Physics, Vol.581, 
Springer-Verlag, Berlin, 2001, p. 195; and refs. therein.\vs
\bibitem{finelli} P. Finelli, N. Kaiser, D. Vretenar and W. Weise, 
\textit{Nucl. Phys.} \textbf{A770} (2006) 1.\vs
\bibitem{lesinski} T. Lesinski, T. Duguet, K. Bennaceur and J. Meyer, 
\textit{Eur. Phys. J.} \textbf{A40} (2009) 121.\vs 
\bibitem{drut} J.E. Drut, R.J. Furnstahl and L. Platter, {\it Prog. Part. 
Nucl. Phys.} {\bf 64} (2010) 120; nucl-th/0906.1463.\vs
\bibitem{platter} S.K. Bogner, R.J. Furnstahl and L. Platter, \textit{Eur. 
Phys. J.} \textbf{A39} (2009) 219.\vs
\bibitem{achim} S.K. Bogner, R.J. Furnstahl, A. Nogga and A. Schwenk, {\it Nucl.
Phys.} {\bf A763} (2005) 59; nucl-th/0903.3366.\vs
\bibitem{roth} R. Roth, P. Papakonstantinou, N. Paar, H. Hergert, T. Neff and
H. Feldmeier, {\it Phys. Rev.} {\bf C73} (2006) 044312.\vs
\bibitem{negele} J.W. Negele and D. Vautherin, {\it Phys. Rev.} {\bf C5} (1972)
1472.\vs 
\bibitem{efun} N. Kaiser, S. Fritsch and W. Weise, \textit{Nucl. Phys.} 
\textbf{A724} (2003) 47.\vs
\bibitem{deltamat} S. Fritsch, N. Kaiser and W. Weise, \textit{Nucl. Phys.} 
\textbf{A750} (2005) 259.\vs
\bibitem{dmeimprov} B. Gebremariam, T. Duguet and S.K. Bogner, 
nucl-th/0910.4979.\vs 
\bibitem{ring} P. Ring and P. Schuck, "The Nuclear Many-Body Problem", Springer
Verlag, (1980); chapters 4 and 5.\vs 
\bibitem{spectral} N. Kaiser, S. Gerstend\"orfer and W. Weise, \textit{Nucl. 
Phys.} \textbf{A637} (1998) 395.\vs
\bibitem{nnpap} N. Kaiser, R. Brockmann and W. Weise, \textit{Nucl. Phys.} 
\textbf{A625} (1997) 758.\vs
\bibitem{fujita} J. Fujita and H. Miyawawa,  {\it Prog. Theor. Phys.} {\bf 17} 
(1957) 360; 366.\vs
\bibitem{quasi} N. Kaiser,  {\it Nucl. Phys.} {\bf A768} (2006) 99.\vs
\bibitem{short} N. Kaiser,  {\it Phys. Rev.} {\bf C70} (2004) 034307.\vs
\bibitem{tueb} O. Plohl and C. Fuchs,  {\it Phys. Rev.} {\bf C74} (2006) 
034325.\vs
\bibitem{note} N. Kaiser and W. Weise,  {\it Nucl. Phys.} {\bf A804} (2008) 
60.\vs
\bibitem{3bodyso} N. Kaiser,  {\it Phys. Rev.} {\bf C68} (2003) 054001.\vs
\end{thebibliography}
\end{document}